\documentclass[a4paper,12pt]{article}

\usepackage{amsmath}
\usepackage[psamsfonts]{amssymb}
\usepackage{rsfs}
\usepackage{bm}

\usepackage{cite}
\usepackage[dvips]{graphicx}
\usepackage{color}

\makeatletter
\@addtoreset{equation}{section}
\makeatother

\begin{document} 

\begin{titlepage}

\baselineskip 10pt
\hrule 
\vskip 5pt
\leftline{}
\leftline{Chiba Univ. Preprint
          \hfill   \small \hbox{\bf CHIBA-EP-177}}
\leftline{\hfill   \small \hbox{July 2009}}
\vskip 5pt
\baselineskip 14pt
\hrule 
\centerline{\Large\bf 
} 
\vskip 0.2cm
\centerline{\Large\bf  
Infrared behavior of the ghost propagator 
}
\vskip 0.2cm
\centerline{\Large\bf  
in the  Landau gauge Yang-Mills theory 
}
\vskip 0.2cm

\vskip 0.2cm

\centerline{{\bf 
Kei-Ichi Kondo,$^{\dagger,{1},{2}}$  
}}  
\vskip 0.2cm
\centerline{\it
${}^{1}$Department of Physics, University of Tokyo,
Tokyo 113-0033, Japan
}
\centerline{\it
${}^{2}$Department of Physics, 
Chiba University, Chiba 263-8522, Japan
}
\vskip 0.2cm

\begin{abstract}
We prove that the Faddeev-Popov ghost dressing function in the Yang-Mills theory is  non-zero and finite in the  limit of vanishing momenta and hence the ghost propagator behaves like free in the deep infrared regime, within the Gribov-Zwanziger framework of the $D$-dimensional $SU(N)$ Yang-Mills theory in the Landau gauge  for any $D>2$. 
This result implies that the Kugo-Ojima color confinement criterion is not satisfied in its original form. 
We point out that the result crucially depends on the explicit form of the non-local horizon term adopted. 
The original Gribov prediction in the Landau gauge should be reconsidered in connection with color confinement. \end{abstract}

 

\hrule  
${}^\dagger$ 
On sabbatical leave of absence from Chiba University. 
\\
 E-mail:  {\tt kondok@faculty.chiba-u.jp}

\par 
\par\noindent


\vskip 0.3cm

\newpage
\pagenumbering{roman}
\tableofcontents




\end{titlepage}


\pagenumbering{arabic}

\baselineskip 14pt
\section{Introduction and main results}

The behaviors of the gluon and ghost propagators in the deep infrared regime is important for clarifying the dynamics of non-perturbative nature in non-Abelian gauge theories.  In particular, it is believed that they are intimately related to quark, gluon and more generally color confinement \cite{KO79,Kugo95}. 

From the viewpoint of the analytical treatment of the quantum field theory, the exact path integration of the quantum Yang-Mills theory in the continuum spacetime is still difficult in the sense that the fundamental modular region is not yet specified in a manageable form of the Lagrangian or Hamiltonian field theory in the Lorentz covariant gauge. 
Nevertheless, it will be important to consider how to incorporate the existence of the Gribov horizon as the boundary of the 1st Gribov region to remove the Gribov copies as much as possible\cite{Gribov78,Zwanziger89,Zwanziger92,Zwanziger93}.

In a previous paper  \cite{Kondo09a}, we have studied an effect on the ghost propagator or ghost dressing function $G(k^2)$ of the horizon condition which plays the role of restricting the functional integral region of the gluon field to the 1st Gribov region. 
We have rewritten the Zwanziger horizon condition in terms of the ghost dressing function and the Kugo-Ojima parameter $u(0)$ for color confinement.  This has enabled one to study which value of the Kugo-Ojima parameter $u(0)$ is allowed if the horizon condition is imposed. 
Although all the calculations were performed in the limit of vanishing Gribov parameter $\gamma$ for simplicity, the obtained value is consistent with the result of numerical simulations.  In fact, the direct measurements on a lattice \cite{FN07,Sternbeck06} show $u(0) = -0.6 \sim -0.8$.   
 Consequently, the ghost propagator behaves like free and the gluon propagator is non-vanishing at low momenta.

In this paper, we study the same issue in  the $D$-dimensional $SU(N)$ Yang-Mills theory in the Landau gauge within the Gribov-Zwanziger framework with a non-zero Gribov parameter $\gamma \ne 0$. 
We discuss how the restriction of the integration region to the (1st) Gribov region constrains the possible value for the Kugo-Ojima parameter for color confinement and the ghost  dressing function.

We prove that the ghost dressing function $G(k^2)$ is  non-zero and finite in the deep IR limit $k \rightarrow 0$ irrespective of the number of color $N$, and hence the ghost propagator behaves like free in the deep infrared regime for $D>2$. 
In addition, with an additional input, we have
\begin{equation}
 G(k^2\rightarrow 0) \rightarrow 3 \quad  (D=4). 
\end{equation}
This result is equivalent to say that the Kugo-Ojima color confinement criterion $u(0)=-1$ is not satisfied in its naive form. 
Rather, we find the exact value for the KO parameter 
\begin{equation}
u(0)=-2/3 \quad (D=4),
\end{equation}
irrespective of the number of color. 
These results are in harmony with the decoupling solution of the Schwinger-Dyson equation \cite{Boucaudetal08}, recent lattice results \cite{BMMP08,BIMPS09,CM07,CM08,OS08,SS08}, and other approaches \cite{ABP08,Dudaletal08,FMP08}.

\section{General settings}
We consider the Gribov-Zwanziger theory \cite{Zwanziger89} for the $D$-dimensional Euclidean $SU(N)$ Yang-Mills theory in the Landau gauge, which is defined by the partition function:
\begin{equation}
 Z_{\gamma} := \int [d\mathscr{A}] \delta (\partial^\mu \mathscr{A}_\mu) \det K \exp \{ -S_{YM}  - \gamma \int d^D x h(x) \} 
 , 
 \label{GZ1}
\end{equation}
where $S_{YM}$ is the Yang-Mills action, 
\begin{align}
 S_{YM}[\mathscr{A}] :=   \int d^Dx \frac14 \mathscr{F}_{\mu\nu}  \cdot \mathscr{F}_{\mu\nu} ,
 \quad
 \mathscr{F}_{\mu\nu} := \partial_\mu \mathscr{A}_\nu - \partial_\nu \mathscr{A}_\mu +g \mathscr{A}_\mu \times \mathscr{A}_\nu ,
\end{align}
$K$ is the Faddeev-Popov operator associated with the Landau gauge fixing condition $\partial^\mu \mathscr{A}_\mu(x)=0$:
\begin{equation}
 K[\mathscr{A}] :=-\partial_\mu D_\mu[\mathscr{A}]=-\partial_\mu (\partial_\mu+g \mathscr{A}_\mu \times)  
  ,
\end{equation}
 $h(x)=h_{[\mathscr{A}]}(x)$ is the Zwanziger horizon function given by  
 \footnote{
We have changed the convention. In \cite{Kondo09a}, $\gamma$ must be replaced by $-\gamma$ in (1.1) and (2.1).  
In \cite{Kondo09a}, the minus sign in (1.2), (1.4), (3.1a), (3.1b) and (3.1c) should be removed. 
The minus sign should be inserted in (3.2). After these changes, the results are unchanged. 
} 
\begin{equation}
 h(x) 
:=   \int d^Dy gf^{ABC} \mathscr{A}_\mu^{B}(x) (K^{-1})^{CE}(x,y) gf^{AFE} \mathscr{A}_\mu^{F}(y)
 . 
\end{equation}
In addition, the parameter $\gamma$ called the Gribov parameter is determined by solving a gap equation, commonly called the horizon condition:
\begin{equation}
 \langle h(x) \rangle_{\gamma} = (N^2-1)D .
\end{equation}
Here the dot and the cross are defined as 
\begin{equation}
  \mathscr{A} \cdot \mathscr{B} := \mathscr{A}^A \mathscr{B}^A , \quad 
  (\mathscr{A} \times \mathscr{B})^A := f^{ABC} \mathscr{A}^B \mathscr{B}^C 
  ,
\end{equation}
using the structure constant of the gauge group $G=SU(N)$.

It should be remarked that the action corresponding to the partition function (\ref{GZ1}) contains the {\it non-local} horizon term: 
\begin{equation}
 \int d^Dx h(x) 
:=   \int d^Dx \int d^Dy gf^{ABC} \mathscr{A}_\mu^{B}(x) (K^{-1})^{CE}(x,y) gf^{AFE} \mathscr{A}_\mu^{F}(y)
 . 
\end{equation}
In what follows, we assume that the Euclidean version can be continued analytically to the Minkowski version by the Wick rotation. 

Moreover, it has been shown \cite{Zwanziger92,Zwanziger93} that the non-local action (\ref{GZ1}) can be put in an equivalent {\it local} form by introducing a set of complex conjugate commuting variables $\xi, \bar\xi$ and anticommuting ones $\omega, \bar\omega$, which we call the localized Gribov-Zwanziger (GZ) action. 
In fact, the horizon term can be rewritten into the local form by introducing auxiliary fields (See e.g., \cite{Kondo09b}):
\begin{equation}
 e^{ - \gamma \int d^D x h(x) } 
=
\int [d\xi] [d\bar{\xi}]  [d\omega] [d\bar{\omega}]  
 \exp \left\{   - \tilde{S}_\gamma[\mathscr{A},\xi,\bar{\xi},\omega,\bar{\omega}]  \right\} 
 ,
\end{equation}
where
\begin{align}
    \tilde{S}_\gamma[\mathscr{A},\xi,\bar{\xi},\omega,\bar{\omega}]  
=: \int d^Dx & [
  \bar{\xi}_\mu^{CA} K^{AB} \xi_\mu^{CB} 
-  \bar{\omega}_\mu^{CA} K^{AB} \omega_\mu^{CB}
\nonumber\\&
+ i \gamma^{1/2} gf^{ABC} \mathscr{A}_\mu^B  \xi_\mu^{AC} + i \gamma^{1/2} gf^{ABC} \mathscr{A}_\mu^B \bar{\xi}_\mu^{AC} 
 ] 
  .  
  \label{aux-action}
\end{align}
Thus, the localized  Gribov-Zwanziger theory is given by
\begin{equation}
 Z_{\rm GZ}  := \int [d\mathscr{A}] [d\mathscr{B}][d\mathscr{C}][d\bar {\mathscr{C}}] [d\xi] [d\bar{\xi}]  [d\omega] [d\bar{\omega}]  \exp \{ - S_{\rm YM}^{\rm tot} - \tilde{S}_\gamma  \} 
 , 
 \label{GZ2}
\end{equation}
where we have introduced the Nakanishi-Lautrup auxiliary field $\mathscr{B}$, the Faddeev-Popov ghost field $\mathscr{C}$ and the antighost field $\bar{\mathscr{C}}$, 
\begin{align}
   S_{\rm GZ} :=& S_{\rm YM}^{\rm tot}[\mathscr{A},\mathscr{C},\bar{\mathscr{C}},\mathscr{B}] + \tilde{S}_\gamma [\mathscr{A},\xi,\bar{\xi},\omega,\bar{\omega}]  
  \nonumber\\
=& S_{\rm YM}[\mathscr{A}] + S_{\rm GF+FP}[\mathscr{A},\mathscr{C},\bar{\mathscr{C}},\mathscr{B}] 
+ \tilde{S}_\gamma [\mathscr{A},\xi,\bar{\xi},\omega,\bar{\omega}]
 ,
\end{align}
with the gauge fixing and the Faddeev-Popov term of the BRST exact form:
\begin{align}
 S_{\rm GF+FP}  :=& \int d^Dx  \left\{ \mathscr{B} \cdot \partial_\mu \mathscr{A}_\mu 
+i \bar {\mathscr{C}} \cdot \partial_\mu D_\mu \mathscr{C} \right\} 
= - \mbox{\boldmath $\delta$}   \left[ i\bar{\mathscr{C}}  \cdot  \left( \partial_\mu \mathscr{A}_\mu  \right) \right]  
  ,
\end{align}
using the BRST transformation $\mbox{\boldmath $\delta$}$.
The localized GZ theory is renormalizable to all orders of perturbation theory. Hence, the restriction to the (first) Gribov region $\Omega$ makes perfect sense at the quantum level, and finite results are obtained consistent with the renormalization group. 
In the limit $\gamma \to 0$, the integrations over the auxiliary fields cancel and the GZ theory reduces to the usual BRST approach for the Yang-Mills theory:
\begin{equation}
 Z_{\rm YM}  := \int [d\mathscr{A}] [d\mathscr{B}][d\mathscr{C}][d\bar {\mathscr{C}}]    \exp \{ -S_{YM}^{\rm tot}  \} 
 . 
\end{equation}

It is known that the localized horizon term breaks the usual BRST invariance of the GZ theory.  Therefore, the GZ theory is no longer invariant under the usual BRST transformation.  In other words, the BRST symmetry is not the exact symmetry for the GZ theory (unless $\gamma=0$). 
Nevertheless, one can find the modified ``BRST'' symmetry for the GZ theory with nilpotency \cite{Kondo09b} (or without nilpotency \cite{Sorella09}), which reduces to the usual BRST symmetry in the limit $\gamma \to 0$. 
However, the modified ``BRST'' symmetry inevitably becomes non-local even in the localized GZ theory.

It should be noted that the Kugo-Ojima (KO) color confinement criterion was obtained in the framework of the usual BRST quantization for the Faddeev-Popov approach, which corresponds to the $\gamma=0$ case of the above Gribov-Zwanziger formulation.
The usual BRST approach does not take care of the Gribov copy problem. 
Therefore, if one begins to avoid the Gribov copy by restricting the space of gauge field configuration, it may happen that the Kugo-Ojima criterion $u(0)=-1$ based on the usual BRST approach does not necessarily hold.

\section{Horizon function and  ghost propagator}

In what follows, we define the  Fourier transform of the two-point function for composite operators by
\begin{equation}
  \langle \phi_1^A \phi_2^B \rangle_{k} := 
\int d^Dx e^{ik(x-y)} \langle 0| T[\phi_1^A(x) \phi_2^B(y) ] |0 \rangle 
 . 
\end{equation} 
In a previous paper \cite{Kondo09a}, the $\gamma=0$ case has been studied and the following identity has been derived.
The average of the horizon function (assuming the translational invariance) is exactly rewritten in terms of three parameters $G(0)$, $u(0)$ and $w(0)$ defined below:
\begin{align}
 \langle  h(0) \rangle
 \equiv& V_D^{-1} \int d^Dx \langle   h(x) \rangle
\nonumber\\
=& 
 - (N^2-1)  \left\{ (D-1)u(0) + G(0)[u(0)+w(0)] \right\}  
 ,
 \label{id1}
\end{align} 
where we have defined the volume $V_D$ of the $D$-dimensional Euclidean space(time).
This expression given in a previous paper is same in the bare case before the renormalization as
\begin{align}
 \langle  h(0) \rangle
= - (N^2-1)  \left\{ Du(0) +w(0) - G(0)[u(0)+w(0)]^2 \right\}  
 .
 \label{id1b}
\end{align}
Here $G(0)$ is the IR limit $k^2 \to 0$ of the ghost dressing function $G(k^2)$ defined from the ghost propagator $\langle  \mathscr{C}^A  \bar{   \mathscr{C}}^B \rangle_k$ by
\begin{align}
 G(k^2) \delta^{AB}  := -k^2  \langle  \mathscr{C}^A  \bar{   \mathscr{C}}^B \rangle_k  
 ,
\end{align} 
$u(0)$ is the Kugo-Ojima parameter defined by the $k^2 \to 0$ limit of the Kugo-Ojima function $u(k^2)$:
\begin{equation}
  \langle  (D_\mu \mathscr{C})^{A}  (g\mathscr{A}_\nu \times \bar{\mathscr{C}})^{B}  \rangle_{k} 
:= \left( \delta_{\mu\nu} - \frac{k_\mu k_\nu}{k^2} \right) \delta^{AB} u(k^2)   
 , 
 \label{KOf}
\end{equation}
and $w(0)$ is the massless pole residue in 
\begin{equation}
\langle   (g \mathscr{A}_\mu \times \mathscr{C})^A (g \mathscr{A}_\nu \times  \bar{\mathscr{C}})^B \rangle_k^{\rm m1PI}
=  \left[ \delta_{\mu\nu} u(k^2) + \frac{k_\mu k_\nu}{k^2} w(k^2) \right] \delta^{AB} 
 , 
 \label{keyid}
\end{equation}
where the modified one-particle irreducible  (m1PI) part of
$
\langle   (g \mathscr{A}_\mu \times \mathscr{C})^A (g \mathscr{A}_\nu \times  \bar{\mathscr{C}})^B \rangle_k
$ 
is defined by (the diagrammatic representation is give later.)
\begin{align}
& \langle   (g \mathscr{A}_\mu \times \mathscr{C})^A (g \mathscr{A}_\nu \times  \bar{\mathscr{C}})^B \rangle_k
\nonumber\\
:=& 
\langle   (g \mathscr{A}_\mu \times \mathscr{C})^A (g \mathscr{A}_\nu \times  \bar{\mathscr{C}})^B \rangle_k^{\rm m1PI}
+ \langle (g \mathscr{A}_\mu \times \mathscr{C})^A \bar{   \mathscr{C}}^C \rangle_k^{\rm 1PI} 
 \langle \mathscr{C}^C \bar{\mathscr{C}}^D \rangle_k 
 \langle \mathscr{C}^D (g \mathscr{A}_\nu \times \bar{   \mathscr{C}})^B \rangle_k^{\rm 1PI} 
  ,
\label{def3}  
\end{align}
with the IPI part of $\langle (g \mathscr{A}_\mu \times \mathscr{C})^A \bar{   \mathscr{C}}^B \rangle_k$ and $\langle \mathscr{C}^A (g \mathscr{A}_\nu \times \bar{   \mathscr{C}})^B \rangle_k$ defined by
\begin{align}
 \langle (g \mathscr{A}_\mu \times \mathscr{C})^A \bar{\mathscr{C}}^B \rangle_k
:=& 
 \langle (g \mathscr{A}_\mu \times \mathscr{C})^A \bar{   \mathscr{C}}^C \rangle_k^{\rm 1PI} 
\langle   \mathscr{C}^C \bar{\mathscr{C}}^B \rangle_k 
,
\\
\langle   \mathscr{C}^A (g \mathscr{A}_\nu \times  \bar{\mathscr{C}})^B \rangle_k 
 :=&  
 \langle \mathscr{C}^A \bar{\mathscr{C}}^C \rangle_k 
\langle \mathscr{C}^C (g \mathscr{A}_\nu \times \bar{   \mathscr{C}})^B \rangle_k^{\rm 1PI}    
 .
 \label{f2b}
\end{align}

In the Landau gauge, especially, the ghost dressing function $G(k^2)$  is related to the two functions  $u(k^2)$ and $w(k^2)$ as
\begin{align}
 G(k^2)   
 =     [1+u(k^2)+w(k^2)]^{-1}  
 .
 \label{Fid}
\end{align} 
In the Landau gauge, there is a symmetry for the exchange between ghost and antighost, called the Faddeev-Popov conjugation invariance. See Appendix~A.

In the next section, we confirm that all the relations in the above hold also in the  $\gamma \ne 0$ case of the Gribov-Zwanziger theory, although they are originally derived in the $\gamma=0$ case \cite{Kondo09a}.  Therefore, the conclusions derived from them in the previous paper \cite{Kondo09a} are valid also in the  $\gamma \ne 0$ case of the Gribov-Zwanziger theory.  
Although the Gribov-Zwanziger theory involves $\gamma$, these relations do not involve the explicitly $\gamma$-dependent extra terms and the $\gamma$ dependence appears only through the functions $F$, $u$ and $w$ implicitly.

From (\ref{id1}) or (\ref{id1b}), due to the horizon condition $\langle  h(0) \rangle =  (N^2-1)D < +\infty$, it is obvious that $G(0)$ is finite as far as $u(0)$ and $w(0)$ are finite. 
The ghost dressing constant obeys 
\begin{equation}
G(0) 
= - \frac{(D-1)u(0)+\langle  h(0) \rangle/(N^2-1)}{u(0)+w(0)}
= - \frac{(D-1)u(0)+D}{u(0)+w(0)}
 .
 \label{F1}
\end{equation}
On the other hand, we have from   (\ref{Fid})
\begin{align}
 G(0) = [1+u(0)+w(0)]^{-1} 
 .
 \label{F2}
\end{align} 
If $u(0) + w(0) = 0$,  then $G(0)$ is finite, i.e., $G(0)=1$, due to  (\ref{F2}). In this case, we have $u(0)=-w(0)=-D/(D-1)$ from the consistency with (\ref{F1}).
If $u(0) + w(0) \ne 0$, then $G(0)$  can not be divergent and is finite due to the horizon condition 
$\langle  h(0) \rangle =  (N^2-1)D < +\infty$, see (\ref{F1}).
In any case, thus, the ghost dressing function $G(k^2)$ is finite in the deep IR limit.

It was implicitly  assumed in Kugo \cite{Kugo95} that 
\footnote{
The author would like to thank Prof. Taichiro Kugo for correspondence on this issue.
}
\begin{equation}
 w(0) = 0 .
\end{equation}
It should be checked whether $w(0) = 0$ is true or not. 
According to  numerical simulations on a lattice  \cite{Sternbeck06,SIMPS06} (see Figure 5.4 in section 5.2.2 \cite{Sternbeck06}) and an independent study \cite{ABP09} based on \cite{ABPRQ09,GHQ04}, $w(0) = 0$ seems to be true. In this case, 
\begin{equation}
G(0) 
= -D+1+ \frac{D}{-u(0)}
 .
\end{equation}
In the case of $w(0)=0$, therefore, we have two relations:
\begin{subequations}
\begin{align}
& \text{$\bullet$ a relationship} \quad
\langle  h(0) \rangle  
  = - (N^2-1) 
  \left\{ -Du(0) + \frac{u(0)^2}{1+u(0)} \right\}   
\label{r1}
  ,
\\
& \text{$\bullet$ the horizon condition} \quad
 \langle  h(0) \rangle  =  (N^2-1)D .
\end{align}
\end{subequations}
Equating two relations, we obtain an algebraic equation for $u(0)$. 
Consequently, $u(0)$ is determined by solving the algebraic equation as
\begin{equation}
 (D-1)u(0)^2+2Du(0)+D=0  \Longrightarrow
u(0)=(-D \pm \sqrt{D})/(D-1) ,
\end{equation}
which implies
\footnote{If we do not assume $w(0)=0$, we have
\begin{equation}
G(0) = 1 +(1-D)w(0)/2+ \sqrt{[1+(1-D)w(0)/2]^2-1+D} > 0.
\end{equation} 
by solving 
\begin{equation}
  G^{2}(0) -[2+(1-D)w(0)]G(0) + 1-D = 0 . 
\end{equation}
}
\begin{equation}
  G^{2}(0) -2G(0) + 1-D = 0 , \quad G(0) = 1 \pm \sqrt{D}  .
\end{equation} 
Note that $u(k^2)$ cannot be smaller than $-1$, i.e., $u(k^2) \ge -1$  for $G(k^2)$ to be non-negative, $G(k^2) \ge 0$.

For $D=4$,  $3u(0)^2+8u(0)+4=0$ has solutions $u(0)=-2/3$ and $-2$.  We must adopt the solution $u(0)=-2/3$. 
Thus we have  
\begin{equation}
u(0)=- \frac23 \sim -0.66666... , \quad 
G(0) = [1+u(0)]^{-1} = 3 
  ,
\end{equation}
\textit{irrespective of the number of color $N$}. 
Thus the ghost dressing function $G(k^2)$ is finite  even in the deep infrared limit $k^2 \to 0$.
 This value seems to agree with the bare value obtained in numerical simulations on a lattice, see Figure 5 of \cite{BMMP08} for SU(2) and Figure 4 of \cite{BIMPS09}  for SU(3) 
(and Figure 2 of \cite{CM07} for the renormalized values).
The renormalization point dependence will be discussed later.

For $D=3$,  $2u(0)^2+6u(0)+3=0$ has solutions $u(0)=(-3 \pm \sqrt{3})/2$.   We must adopt the solution
\begin{equation}
u(0)= (-3 + \sqrt{3})/2 \sim -0.63397... , \quad 
G(0) = [1+u(0)]^{-1} = 1  + \sqrt{3} = 2.73205... 
\end{equation}
This should be compared with numerical simulations on a lattice, e.g., \cite{CM08}, where the renormalization must be properly taken into account (separately).

For $D=2$, the derived relations are also valid.  However, in two-dimensional field theory, a subtle problem exists. 
As is known in the Coleman theorem for the absence of the Nambu-Goldstone particle in two dimensions, a massless particle cannot be well defined for $D=2$. If the ghost dressing function for $D=2$ is finite, this remark will be applied. 
In fact, the numerical simulations indicate that the ghost dressing function is not finite and the ghost propagator becomes more singular than the free one for $D=2$ \cite{Maas09,CM08}.

The obtained result differs from the old Gribov prediction in the Landau gauge.\footnote{
In the Coulomb gauge, the situation is different and there are at present no contradictions with the Gribov prediction in the Coulomb gauge \cite{Coulomb}. 
}
 It is possible to reconcile this result with the original Gribov argument as follows.  
In order to see where the difference comes from, we expand the horizon function   
\begin{align}
 \gamma h(x) 
:=&   \gamma \int d^Dy gf^{ABC} \mathscr{A}_\mu^{B}(x) \left( \frac{1}{-\partial_\nu D_\nu[\mathscr{A}]} \right)^{CE}(x,y) gf^{AFE} \mathscr{A}_\mu^{F}(y)
\nonumber\\
=&   \gamma \int d^Dy gf^{ABC} \mathscr{A}_\mu^{B}(x) \left( \frac{1}{-\partial_\nu \partial_\nu} \right) \delta^{CE} \delta^{(D)}(x-y) gf^{AFE} \mathscr{A}_\mu^{F}(y) + O((gA)^3)
\nonumber\\
=&   \gamma N  g \mathscr{A}_\mu^{B}(x) \left( \frac{1}{-\partial_\nu \partial_\nu} \right)   g \mathscr{A}_\mu^{B}(x) + O((gA)^3)
 .
\nonumber
\end{align}
Then the quadratic part in the gauge field of the Gribov-Zwanziger Lagrangian is modified as
\begin{equation}
 \mathscr{L}_{\ YM} + \gamma h 
= \mathscr{A}_\mu^{A} \left[  (-\partial^2)   + \gamma N  g^2  \left( \frac{1}{-\partial^2} \right)    \right] \mathscr{A}_\mu^{A}  + O((gA)^3)
 . 
\nonumber
\end{equation}
The resulting gluon propagator up to $O((gA)^2)$ vanishes in the IR limit:
\begin{equation}
  \left[  (-\partial^2) + \gamma N  g^2  \left( \frac{1}{-\partial^2} \right)    \right]^{-1}
  = \frac{-\partial^2}{(-\partial^2)^2+Ng^2 \gamma}
  \Longrightarrow 
  \frac{k^2}{(k^2)^2+Ng^2 \gamma} \downarrow 0 \ (k \downarrow 0)
 . 
\nonumber
\end{equation}
In this way the Gribov prediction is reproduced from the Gribov-Zwanziger theory.
However, we can raise the question: Do the higher order terms modify the IR behavior from the Gribov prediction?
To answer this question, the full or exact estimation of the horizon function is needed as we have done in this paper.

In our result,  the Gribov result was obtained by taking into account only the lowest order term as shown in the above. The formal power series expansion, 
\footnote{
It should be noted that this expansion is not the same as the perturbative expansion in the coupling constant.
Therefore, this result does not necessarily agree with the result of  loop expansions.\cite{Gracey06}
}
$
  \frac{1}{1+u(0)} =  [1+u(0)]^{-1}
  = 1 - u(0) + u(0)^2 + \cdots 
$
yields the horizon condition:
\begin{align}
 \langle  h(0) \rangle
  =&  (N^2-1) 
  \left\{ - Du(0) + u(0)^2 - u(0)^3 + \cdots  \right\}    =(N^2-1)D 
  .
 \label{f23}
\end{align}
If we took into account only a linear term in $u(0)=O(g^2)$, then the horizon condition would lead to the Kugo-Ojima criterion $u(0)=-1$ and the divergent ghost dressing function $G(0) =[1+u(0)]^{-1}=\infty$. 
In this way we can reproduce the Gribov original result in the Landau gauge in the lowest order.
This result totally changes if we include the effect of higher order terms.
Thus, the ghost propagator behaves like free at low momenta. Whereas the gluon propagator is expected to be non-vanishing at low momenta where the gluon dressing function vanishes in the IR limit.  This is possible as shown in \cite{Kondo04}.

\section{Proof}

The proof  of (\ref{id1}) given in \cite{Kondo09a} in the case of  $\gamma=0$  is extended to the case of $\gamma \ne 0$ as follows. 
Although the relations we establish in the following have the same forms as those given in section 2 of \cite{Kondo09a}, the Green (or correlation) functions have the implicit $\gamma$ dependence, and hence the average should be understood to be $\gamma$ dependent  $\langle ... \rangle^{\gamma}$ in the Gribov-Zwanziger theory, which is however omitted for simplicity.

\subsection{The relation (2.5)}
The same relation as (2.5) of \cite{Kondo09a} in the case of  $\gamma \ne 0$:
\begin{equation}
 \langle (D_\mu \mathscr{C})^A \bar{\mathscr{C}}^B \rangle_k
 = i\frac{k_\mu}{k^2} \delta^{AB} 
 \label{DCC}
\end{equation} 
 is derived in the same way as in the $\gamma=0$ case.
 
In the path-integral quantization, it is derived as follows. 
\begin{align}
 & \langle 0|\partial ^{\mu}(D_{\mu}\mathscr{C})^A(x)\bar{\mathscr{C}}^B(y)|0 \rangle 
\nonumber\\
 =& \int d\mu(\Phi)  e^{-S_{\rm GZ}}  \partial ^{\mu}(D_{\mu}\mathscr{C})^A(x) \bar{\mathscr{C}}^B(y)
\nonumber\\
 =& \int d\mu(\Phi)    \frac{\delta e^{-S_{\rm GZ}}}{\delta (-\bar{\mathscr{C}}^A(x))}\bar{\mathscr{C}}^B(y)
\nonumber\\
 =&-\int d\mu(\Phi) \frac{\delta}{\delta\bar{\mathscr{C}}^A(x)}\left[e^{-S_{\rm GZ}}\bar{\mathscr{C}}^B(y)\right]+\int d\mu e^{-S_{\rm GZ}}\frac{\delta \bar{\mathscr{C}}^B(y)}{\delta \bar{\mathscr{C}}^A(x)}
\nonumber\\
 =& \delta ^{AB}\delta ^D(x-y)
  ,
  \label{dDCC}
\end{align}
where we have used a fact that the integration of the derivative is identically zero:
\footnote{
This relation is called the Schwinger-Dyson equation, which is a consequence of 
$
  \int_{a}^{b} d\phi \frac{d f}{d \phi}
  = f(b)-f(a) =0 
$
where we have used $f(b)=f(a)$ at the boundaries.
This relation follows also from the shift invariance of the measure, since 
$
  \int d\phi f(\phi) = \int d(\phi+a) f(\phi+a) = \int d\phi f(\phi+a) 
= \int d\phi f(\phi)+a \int d\phi \frac{df(\phi)}{d\phi} + O(a^2)  
$
holds for arbitrary $a$.
}
\begin{equation}
 \int d\mu(\Phi) \frac{\delta}{\delta\Phi _I(x)}\left[\cdots\right] \equiv 0 .
\end{equation}
Another derivation is given in Appendix~B.

\subsection{The 1PI part}

The localized Gribov-Zwanziger theory has the 6 full  propagators:
\begin{align}
 & \mathscr{A}_\mu^A-\mathscr{A}_\nu^B ,
 \nonumber\\
 & \mathscr{A}_\mu^A-\mathscr{B}^B ,
 \nonumber\\
 & \mathscr{A}_\mu^A-\xi_\nu^{BC} ,
\quad  \mathscr{A}_\mu^A-\bar\xi_\nu^{BC} ,
 \nonumber\\
 & \mathscr{C}^A-\bar{\mathscr{C}}^B ,
 \nonumber\\
 & \xi_\mu^{AB}-\bar\xi_\nu^{EF} ,
 \nonumber\\
 & \omega_\mu^{AB}-\bar\omega_\nu^{EF}
 ,
\end{align}
and 5 full vertices:
\begin{align}
 & \mathscr{A}_\mu^A-\mathscr{A}_\rho^B-\mathscr{A}_\sigma^C ,
 \nonumber\\
 & \mathscr{A}_\mu^A-\mathscr{A}_\nu^B-\mathscr{A}_\rho^C-\mathscr{A}_\sigma^D ,
 \nonumber\\
 & \mathscr{A}_\mu^A-\mathscr{C}^B-\bar{\mathscr{C}}^C ,
 \nonumber\\
 & \mathscr{A}_\mu^A-\xi_\rho^{BC}-\bar\xi_\sigma^{EF} ,
 \nonumber\\
 & \mathscr{A}_\mu^A-\omega_\rho^{BC}-\bar\omega_\sigma^{EF}
 ,
\end{align}
where in the tree level only the propagators $\langle \mathscr{A}_\mu^A(x)\xi_\nu^{BC}(y) \rangle$ and $\langle \mathscr{A}_\mu^A(x)\bar\xi_\nu^{BC}(y) \rangle$ have the $\gamma$ dependence. 
We denote a set of all fields by $\Phi:=\{ \mathscr{A}, \mathscr{B}, \mathscr{C}, \bar{\mathscr{C}}, \xi, \bar\xi, \omega, \bar\omega \}$.

\begin{figure}[ptb]
\begin{center}
\includegraphics[width=5.5in]{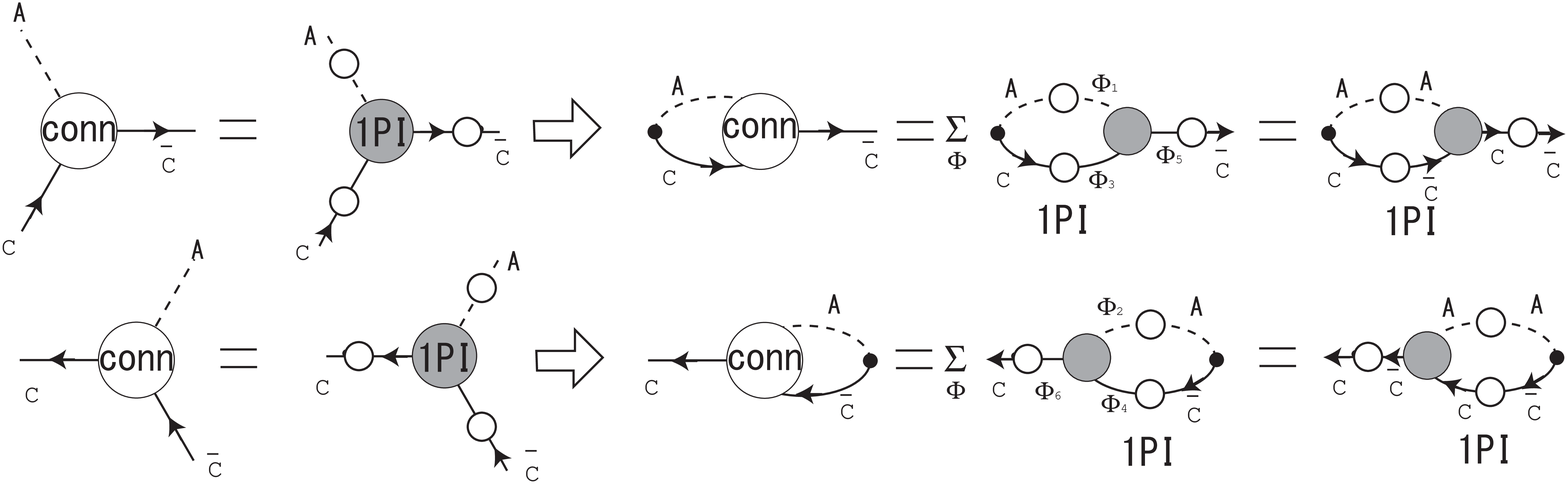}
\end{center}
\caption{Diagrammatic representation of 
(a) $\langle (g \mathscr{A}_\mu \times \mathscr{C})^A \bar{   \mathscr{C}}^B \rangle_k$ and
 $\langle (g \mathscr{A}_\mu \times \mathscr{C})^A \bar{   \mathscr{C}}^B \rangle_k^{\rm 1PI}$, 
(b)  $\langle \mathscr{C}^A (g \mathscr{A}_\nu \times \bar{   \mathscr{C}})^B \rangle_k$ and
$\langle \mathscr{C}^A (g \mathscr{A}_\nu \times \bar{   \mathscr{C}})^B \rangle_k^{\rm 1PI}$.
}
\label{fig:3p-diagram}
\end{figure}

We consider the definition (2.6) and (2.7) for the 1PI part of $\langle (g \mathscr{A}_\mu \times \mathscr{C})^A \bar{   \mathscr{C}}^B \rangle_k$ and $\langle \mathscr{C}^A (g \mathscr{A}_\nu \times \bar{   \mathscr{C}})^B \rangle_k$.
For the 3-point function of gluon, ghost and antighost, there is no disconnected part. 
See Fig.~\ref{fig:3p-diagram} for a diagrammatic representation. 
In defining the 1PI part, the possible intermediate fields $\Phi_1, \Phi_3, \Phi_5$ are uniquely determined to be $\Phi_3=\bar{\mathscr{C}}$, $\Phi_5=\mathscr{C}$ and $\Phi_1=\mathscr{A}$ by taking into account the propagators and vertices enumerated in the above. 
Therefore, the 1PI part is immediately defined by
\begin{align}
 \langle (g \mathscr{A}_\mu(x) \times \mathscr{C}(x))^A \bar{   \mathscr{C}}^B(y) \rangle 
=& \langle (g \mathscr{A}_\mu(x) \times \mathscr{C}(x))^A \bar{\mathscr{C}}^C(z) \rangle^{\rm 1PI} 
 \langle   \mathscr{C}^C(z) \bar{\mathscr{C}}^B(y) \rangle 
  .
\end{align}
In the similar way, we can define
\begin{align}
 \langle \mathscr{C}^A(x) (g \mathscr{A}_\nu(y) \times \bar{   \mathscr{C}}(y))^B \rangle  
=& \langle   \mathscr{C}^A(x) \bar{\mathscr{C}}^C(z) \rangle   
\langle   \mathscr{C}^C(z) (g \mathscr{A}_\nu(y) \times  \bar{\mathscr{C}}(y))^B \rangle^{\rm 1PI} 
  .
\end{align}
Then the Fourier transform reads
\begin{align}
 \langle (g \mathscr{A}_\mu \times \mathscr{C})^A \bar{\mathscr{C}}^B \rangle_k
=& 
 \langle (g \mathscr{A}_\mu \times \mathscr{C})^A \bar{   \mathscr{C}}^C \rangle_k^{\rm 1PI} 
\langle   \mathscr{C}^C \bar{\mathscr{C}}^B \rangle_k 
,
\\
\langle   \mathscr{C}^A (g \mathscr{A}_\nu \times  \bar{\mathscr{C}})^B \rangle_k 
 =&  
 \langle \mathscr{C}^A \bar{\mathscr{C}}^C \rangle_k 
\langle \mathscr{C}^C (g \mathscr{A}_\nu \times \bar{   \mathscr{C}})^B \rangle_k^{\rm 1PI}    
 .
 \label{f2bb}
\end{align}


\subsection{The identity (2.10) or (2.11)}

An identity  (2.10) or (2.11) of \cite{Kondo09a}:
\begin{align}
 0 = \langle   (\partial^\mu D_\mu \mathscr{C})^A (g \mathscr{A}_\nu \times  \bar{\mathscr{C}})^B \rangle_k
 \label{f3}
\end{align}
 is necessary to define the Kugo-Ojima function according to (\ref{KOf}). 
This is derived in the case of  $\gamma \ne 0$ in the same way as in the $\gamma=0$ case. 
 The transversality follows from the identity:
\begin{align}
 0 =& \int d\mu(\Phi) \frac{\delta}{\delta \bar{\mathscr{C}}^A(x)} [ e^{-S_{\rm GZ}}(g\mathscr{A}_\nu \times \bar{\mathscr{C}})^B(y) ]
 \nonumber\\
 =& \int d\mu(\Phi) e^{-S_{\rm GZ}} [gf^{ABC} \mathscr{A}_\nu^C(y) \delta^D(x-y)-  \frac{\partial S_{\rm GZ}}{\partial \bar{\mathscr{C}}^A(x)} (g\mathscr{A}_\nu \times \bar{\mathscr{C}})^B(y)]
 \nonumber\\
 =& \int d\mu(\Phi) e^{-S_{\rm GZ}} [gf^{ABC} \mathscr{A}_\nu^C(y) \delta^D(x-y)- \partial^\mu (D_\mu \mathscr{C})^A(x) (g\mathscr{A}_\nu \times \bar{\mathscr{C}})^B(y)]
  ,
\end{align}
which leads to
\begin{align}
 \langle 0| \partial^\mu (D_\mu \mathscr{C})^A(x) (g\mathscr{A}_\nu \times \bar{\mathscr{C}})^B(y) |0 \rangle
 =  gf^{ABC} \langle 0|  \mathscr{A}_\nu^C(y)  |0 \rangle \delta^D(x-y) = 0 
  ,
\end{align}
since $\langle 0|  \mathscr{A}_\nu^C(y)  |0 \rangle=0$ from the Lorentz invariance of the vacuum.

\subsection{The relations (2.13)--(2.15) and the 1PI part}

We proceed to check the relations (2.13)--(2.15) of \cite{Kondo09a}.

 In order to examine the constraint coming from the horizon condition in the Gribov-Zwanziger theory, the most non-trivial issue in discriminating between $\gamma=0$ and $\gamma \ne 0$ cases is how to connect the function $\langle   (g \mathscr{A}_\mu \times \mathscr{C})^A (g \mathscr{A}_\mu \times  \bar{\mathscr{C}})^A \rangle_{k}$ in the average of the horizon function $\langle  h(0) \rangle$
to the 1PI part\footnote{ 
Here we have used the identity which holds for any functional $ f(\mathcal{A})$ of $\mathcal{A}$:
\begin{equation}
 \langle  f(\mathcal{A})  \mathscr{C}^{A}(x)  \bar{\mathscr{C}}^{B}(y)  \rangle 
=   - \langle f(\mathcal{A}) (K^{-1})^{AB}(x,y) \rangle 
 . 
\nonumber
\end{equation} 
This is consistent with (\ref{dDCC}).
}
\begin{align}
 \langle  h(0) \rangle
 \equiv  V_D^{-1} \int d^Dx \langle   h(x) \rangle
=  - \lim_{k  \rightarrow 0} \langle   (g \mathscr{A}_\mu \times \mathscr{C})^A (g \mathscr{A}_\mu \times  \bar{\mathscr{C}})^A \rangle_{k}
 .
\end{align}

In what follows, we check that no $\gamma$-dependent extra terms do appear in this process even in the Gribov-Zwanziger theory. As a result, the relations derived in the previous paper \cite{Kondo09a} still hold with no change even in the case of $\gamma \ne 0$.  In other words, the relations and the results derived from them in the previous paper \cite{Kondo09a} are valid irrespective of the value of $\gamma$.


\begin{figure}[ptb]
\begin{center}
\includegraphics[width=5.0in]{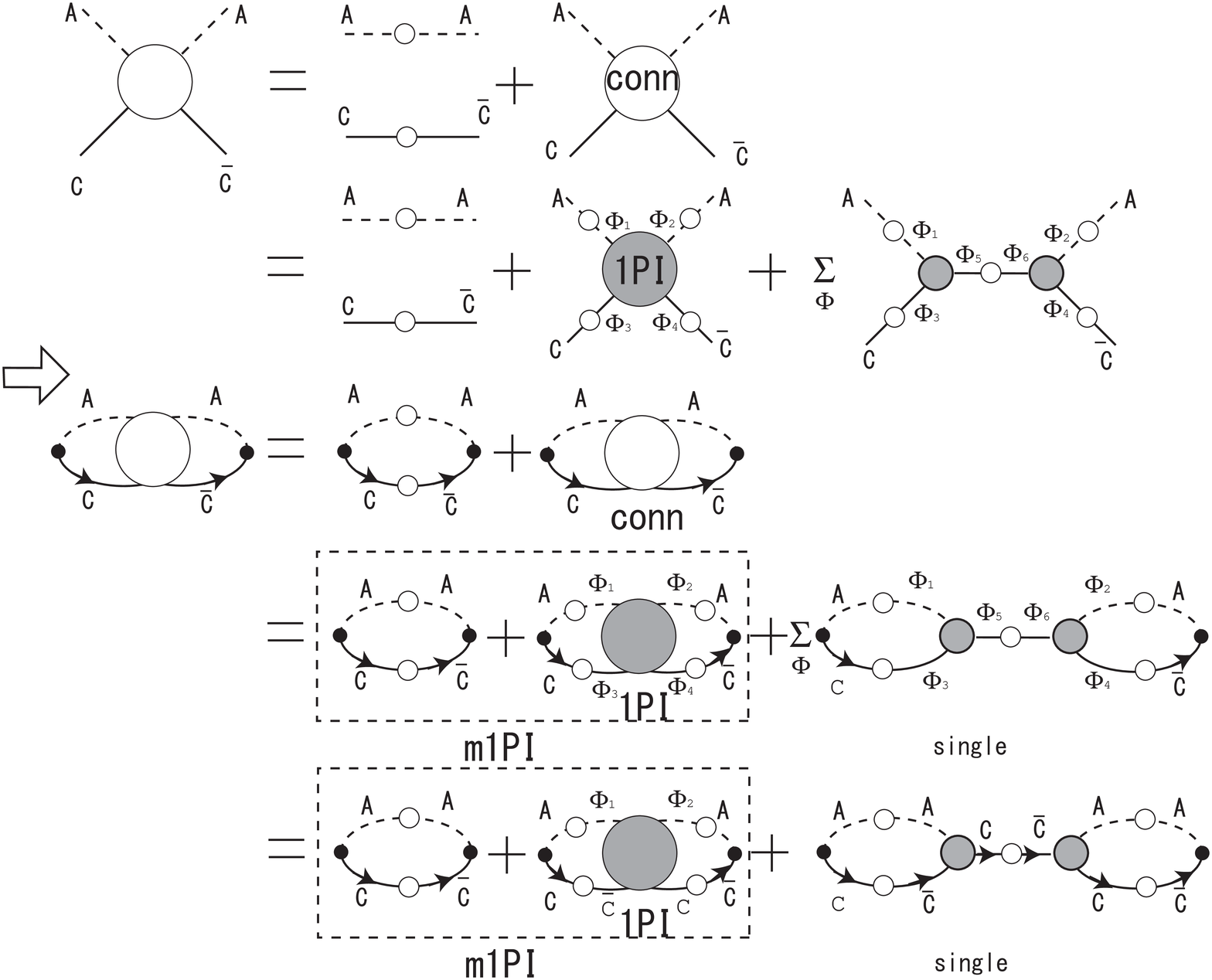}
\end{center}
\caption{Diagrammatic representation of 
$\langle   (g \mathscr{A}_\mu \times \mathscr{C})^A (g \mathscr{A}_\nu \times  \bar{\mathscr{C}})^B \rangle_k$,
$\langle   (g \mathscr{A}_\mu \times \mathscr{C})^A (g \mathscr{A}_\nu \times  \bar{\mathscr{C}})^B \rangle_k^{\rm conn}$, 
$\langle   (g \mathscr{A}_\mu \times \mathscr{C})^A (g \mathscr{A}_\nu \times  \bar{\mathscr{C}})^B \rangle_k^{\rm 1PI}$
and 
$\langle   (g \mathscr{A}_\mu \times \mathscr{C})^A (g \mathscr{A}_\nu \times  \bar{\mathscr{C}})^B \rangle_k^{\rm m1PI}$.
}
\label{fig:4p-diagram}
\end{figure}


The 4-point function of gluon, gluon, ghost and antighost is separated into the connected part and the disconnected part, see Fig.~\ref{fig:4p-diagram}:
\begin{align}
  & \langle   (g \mathscr{A}_\mu(x_1) \times \mathscr{C}(x_2))^A (g \mathscr{A}_\nu(y_1) \times  \bar{\mathscr{C}}(y_2))^B \rangle 
\nonumber\\
:=& gf^{AFC} gf^{BGE} \langle  \mathscr{A}_\mu^F(x_1)   \mathscr{A}_\nu^G(y_1) \rangle 
\langle   \mathscr{C}^C(x_2)   \bar{\mathscr{C}}^E(y_2) \rangle
\nonumber\\
&+ 
 \langle   (g \mathscr{A}_\mu(x_1) \times \mathscr{C}(x_2))^A (g \mathscr{A}_\nu(y_1) \times  \bar{\mathscr{C}}(y_2))^B \rangle^{\rm conn}  
  .
\end{align}
After putting $x_1=x_2:=x$ and $y_1=y_2:=y$ and taking the Fourier transform, we obtain the necessary relations. 
The connected part is defined by
\footnote{
Eq.(2.15) in \cite{Kondo09a} involves misprints in the indices. Here we give correct expressions.
}
\begin{align}
&  \langle   (g \mathscr{A}_\mu \times \mathscr{C})^A (g \mathscr{A}_\nu \times  \bar{\mathscr{C}})^B \rangle_k
\nonumber\\
:=& 
 \int \frac{d^D p}{(2\pi)^D} gf^{AFC} gf^{BGE} \langle  \mathscr{A}_\mu^F   \mathscr{A}_\nu^G \rangle_{k+p}
\langle   \mathscr{C}^C   \bar{\mathscr{C}}^E \rangle_p
 + \langle   (g \mathscr{A}_\mu \times \mathscr{C})^A (g \mathscr{A}_\nu \times  \bar{\mathscr{C}})^B \rangle_k^{\rm conn}
 .
\end{align}

The 1PI part is defined from the connected part as shown in Fig.~\ref{fig:4p-diagram} by using a diagrammatic representation. 
Here we use the same type of argument to define the 1PI part. 
Consequently, the 1PI part is defined from the connected part
\begin{align}
   \langle   (g \mathscr{A}_\mu \times \mathscr{C})^A (g \mathscr{A}_\nu \times  \bar{\mathscr{C}})^B \rangle_k^{\rm conn}
=   \langle   (g \mathscr{A}_\mu \times \mathscr{C})^A (g \mathscr{A}_\nu \times  \bar{\mathscr{C}})^B \rangle_k^{\rm 1PI}
 +    \Delta_{\mu\nu}^{AB}(k) 
 ,
\end{align}
where $\Delta_{\mu\nu}^{AB}$ involves the single ghost propagator:
\begin{align}
 \Delta_{\mu\nu}^{AB}(k)   :=  \langle (g \mathscr{A}_\mu \times \mathscr{C})^A \bar{   \mathscr{C}}^C \rangle_k^{\rm 1PI} 
 \langle \mathscr{C}^C \bar{\mathscr{C}}^D \rangle_k 
 \langle \mathscr{C}^D (g \mathscr{A}_\nu \times \bar{   \mathscr{C}})^B \rangle_k^{\rm 1PI}   
   .
\label{single}  
\end{align}
Therefore, 1PI and m1PI is related as 
\begin{align}
\lambda_{\mu\nu}^{AB}(k) :=&  \langle   (g \mathscr{A}_\mu \times \mathscr{C})^A (g \mathscr{A}_\nu \times  \bar{\mathscr{C}})^B \rangle_k^{\rm m1PI}
\nonumber\\
:=& \int \frac{d^D p}{(2\pi)^D} gf^{AFC} gf^{BGE} \langle  \mathscr{A}_\mu^F   \mathscr{A}_\nu^G \rangle_{k+p}
\langle   \mathscr{C}^C   \bar{\mathscr{C}}^E \rangle_p
  + \langle   (g \mathscr{A}_\mu \times \mathscr{C})^A (g \mathscr{A}_\nu \times  \bar{\mathscr{C}})^B \rangle_k^{\rm 1PI}
   .
   \label{def-lambda}
\end{align}
Then, we can write the relation
\begin{align}
   \langle   (g \mathscr{A}_\mu \times \mathscr{C})^A (g \mathscr{A}_\nu \times  \bar{\mathscr{C}})^B \rangle_k
=    \lambda_{\mu\nu}^{AB}(k)
+  \Delta_{\mu\nu}^{AB}(k)
 .
\end{align}
Thus the average of the horizon function is written as
\begin{align}
 \langle  h(0) \rangle
=  - \lim_{k^2 \rightarrow 0} \langle   (g \mathscr{A}_\mu \times \mathscr{C})^A (g \mathscr{A}_\mu \times  \bar{\mathscr{C}})^A \rangle_{k}
=  - \lambda_{\mu\mu}^{AA}(0) -  \Delta_{\mu\mu}^{AA}(0)
 .
\end{align}

\subsection{The relations (2.16)--(2.25)}

The relations 
(2.16)--(2.25) of \cite{Kondo09a} remain unchanged. 

In the manifestly covariant gauge of the Lorenz type, an idenity holds ((2.17) of \cite{Kondo09a})
\begin{align}
 ik_\mu  \lambda_{\mu\nu}^{AB}(k)
=   \langle \mathscr{C}^A (g \mathscr{A}_\nu \times \bar{   \mathscr{C}})^B \rangle_k^{\rm 1PI}  
 .
 \label{3pa}
\end{align} 
In the Landau gauge, a similar identity also hold ((4.1) of \cite{Kondo09a})
\begin{align}
  \lambda_{\mu\nu}^{AB}(k) (-ik_\nu)
= \langle (g \mathscr{A}_\mu \times \mathscr{C})^A  \bar{   \mathscr{C}}^B \rangle_k^{\rm 1PI}  
 .
 \label{3pb}
\end{align} 
This identity (\ref{3pb}) is derived from (\ref{3pa}) using the Faddeev-Popov conjugation invariance in the Landau gauge. See Appendix~A.

In addition, an identity holds in the manifestly covariant gauge, ((2.3) or (2.8) of \cite{Kondo09a})
\begin{equation}
 \langle (g \mathscr{A}_\mu \times \mathscr{C})^A \bar{   \mathscr{C}}^B \rangle_k^{\rm 1PI} 
= -i k_\mu \left( - \delta^{AB} + \frac{-1}{k^2} \langle   \mathscr{C}^A \bar{\mathscr{C}}^B \rangle_k^{-1}   \right) 
 ,
\label{f2}
\end{equation}
yielding    
\begin{subequations}
\begin{equation}
 i k_\mu \langle (g \mathscr{A}_\mu \times \mathscr{C})^A \bar{   \mathscr{C}}^B \rangle_k^{\rm 1PI} 
=  - \delta^{AB} k^2  - \langle   \mathscr{C}^A \bar{\mathscr{C}}^B \rangle_k^{-1}   
 ,
\label{SD2}
\end{equation}
which is the same as the Schwinger-Dyson equation for the ghost propagator  ((2.9) of \cite{Kondo09a}):
\begin{equation}
   \langle   \mathscr{C}^A \bar{\mathscr{C}}^B \rangle_k = - \delta^{AB}\frac{1}{k^2}
-i \frac{k^\mu}{k^2} \langle (g \mathscr{A}_\mu \times \mathscr{C})^A \bar{   \mathscr{C}}^B \rangle_k 
 .
\label{SD1}
\end{equation} 
\end{subequations}

Using the general form for the uncontracted 2-point function $\lambda_{\mu\nu}^{AB}(k)$ ((2.25) of \cite{Kondo09a}):
\begin{equation}
\lambda_{\mu\nu}^{AB}(k)
=  \left[ \delta_{\mu\nu} u(k^2) + \frac{k_\mu k_\nu}{k^2} w(k^2) \right] \delta^{AB} 
 , 
 \label{key-id}
\end{equation}
we obtain a relationship between the Kugo-Ojima function and the ghost propagator:
\begin{align}
 G(k^2) \delta^{AB} := -k^2  \langle  \mathscr{C}^A  \bar{   \mathscr{C}}^B \rangle_k  
 =     [1+u(k^2)+w(k^2)]^{-1} \delta^{AB}
 .
 \label{F}
\end{align} 
Incidentally, the relation (\ref{key-id}) is obtained from (\ref{KOf}) and ((2.20) of \cite{Kondo09a})
\begin{align}
   \langle   (D_\mu \mathscr{C})^A (g \mathscr{A}_\nu \times  \bar{\mathscr{C}})^B \rangle_k 
=  \left( \delta_{\mu}{}_{\rho} - \frac{k_\mu k_\rho}{k^2} \right) 
\langle   (g \mathscr{A}_\rho \times \mathscr{C})^A (g \mathscr{A}_\nu \times  \bar{\mathscr{C}})^B \rangle_k^{\rm m1PI} 
  .
 \label{f9}
\end{align}  
Here $w(k^2)$ is an unknown function. 
By combining (\ref{key-id}) with (\ref{3pa}), (\ref{3pb}) and (\ref{F}), we obtain
\begin{align}
    \lambda_{\mu\mu}^{AA}(k)
    =& (N^2-1) [Du(k^2)+w(k^2)] ,
\\
 \Delta_{\mu\mu}^{AA}(k)
 =&  -i \lambda_{\mu\sigma}^{AC}(k) k_\sigma \frac{-G(k^2)}{k^2}\delta^{CD} ik_\rho \lambda_{\rho\mu}^{DA}(k)  
 \nonumber\\
 =& -(N^2-1) G(k^2) [u(k^2)+w(k^2)]^2 
 \nonumber\\
 =&  - (N^2-1)  \frac{[u(k^2)+w(k^2)]^2}{1+u(k^2)+w(k^2)}  
 .
 \label{result}
\end{align}
The existence of the last term $\Delta_{\mu\mu}^{AA}(0)$ is crucial to obtain a finite ghost dressing function, see (\ref{Zwanziger-result}).

\section{Renormalization}

In order to compare our result with the numerical simulations on a lattice and the Schwinger-Dyson equations, we need to consider the renormalization, especially, the dependence of the renormalization point under a given renormalization condition.

We denote the gluon propagator in the Landau gauge by
\begin{equation}
 D_{\mu\nu}^{AB}(k) = \delta^{AB}   \left(\delta_{\mu\nu}-\frac{k_\mu k_\nu}{k^2}  \right) \frac{F(k^2)}{k^2}  
 , 
\end{equation}
and the ghost propagator by
\begin{equation}
 G^{AB}(k) = - \delta^{AB} \frac{G(k^2)}{k^2}  
 . 
\end{equation}
The gluon-ghost-antighost vertex is denoted by 
\begin{equation}
 \Gamma_\nu^{ABC}(p,k) = f^{ABC}\Gamma_\nu(p,k) 
 . 
\end{equation}
In the bare case, $d(k^2)=1/k^2$,  $G(k^2)=1/k^2$ and 
$\Gamma_\nu(p,k)=-k_\nu$.

By using the identity (\ref{keyid}),
\begin{equation}
\lambda_{\mu\nu}^{AB}(k)
:= \langle   (g \mathscr{A}_\mu \times \mathscr{C})^A (g \mathscr{A}_\nu \times  \bar{\mathscr{C}})^B \rangle_k^{\rm m1PI}
=  \left[ \delta_{\mu\nu} u(k^2) + \frac{k_\mu k_\nu}{k^2} w(k^2) \right] \delta^{AB} 
 , 
\end{equation}
 it is easy to show that the functions $u(k^2)$ and $w(k^2)$  obey the equations:
\begin{subequations}
\begin{align}
 u(k^2) =& \frac{1}{(D-1)(N^2-1)} \left[ \lambda_{\mu\mu}^{AA}(k) - \frac{k^\mu k^\nu}{k^2} \lambda_{\mu\nu}^{AA}(k) \right]
  ,
  \label{eq-u}
  \\
 w(k^2) =& \frac{1}{(D-1)(N^2-1)} \left[ D \frac{k^\mu k^\nu}{k^2} \lambda_{\mu\nu}^{AA}(k) - \lambda_{\mu\mu}^{AA}(k)   \right]
  .
  \label{eq-w}
\end{align}
\end{subequations}
The momentum dependence of the functions $u(k^2)$ and $w(k^2)$ is determined from the knowledge of the function:
$
\lambda_{\mu\nu}^{AB}(k)
:=   \langle   (g \mathscr{A}_\mu \times \mathscr{C})^A (g \mathscr{A}_\nu \times  \bar{\mathscr{C}})^B \rangle_{k}^{\rm m1PI}
 ,
$
see Fig.~\ref{fig:4p-diagram}.
This is performed e.g., by substituting the solutions of the Schwinger-Dyson equation for the propagators and the vertex functions into the right-hand sides of (\ref{eq-u}) and (\ref{eq-w}), as done in \cite{ABP09}. 

We consider the multiplicative renormalization in which the renormalization constants are introduced in the standard way:
\begin{subequations}
\begin{align}
  F_R(k^2, \mu^2) 
=& Z_A^{-1}(\mu^2,\Lambda^2) F(k^2, \Lambda^2)
,
\\
  G_R(k^2, \mu^2) 
=&   Z_C^{-1}(\mu^2,\Lambda^2) G(k^2, \Lambda^2)
,
\\
  \Gamma_R^\nu(k, p, \mu^2) 
=&   \tilde{Z}_1(\mu^2,\Lambda^2) \Gamma^\nu(k, p, \Lambda^2)
,
\\
  g_R(\mu^2) 
=&   Z_g^{-1}(\mu^2,\Lambda^2) g(\mu^2)
 ,
\end{align}
\end{subequations}
where $\mu$ is the renormalization point, $\Lambda$ is an ultraviolet cutoff, and 
$Z_g=Z_A^{-1/2} Z_C^{-1} \tilde{Z}_1$.

We carry out the renormalization in such a way as to preserve the identity (\ref{Fid}). 
In order to preserve the identity (\ref{Fid}) after renormalization, we impose that 
\begin{subequations}
\begin{align}
  G_R^{-1}(k^2, \mu^2) 
=& Z_C(\mu^2,\Lambda^2) G^{-1}(k^2, \Lambda^2)
,
\\
  1+u_R(k^2, \mu^2) 
=&   Z_C(\mu^2,\Lambda^2) [1+u(k^2, \Lambda^2)]
,
\\
  w_R(k^2, \mu^2) 
=&   Z_C(\mu^2,\Lambda^2) w(k^2, \Lambda^2)
,
\end{align}
\end{subequations}

Consequently, the Kugo-Ojima function $u_R(k^2)$ is not a renormalization-group invariant quantity. Hence, it is not a  renormalization point $\mu$-independent quantity and the Kugo-Ojima parameter $u_R(0)$ depends on the renormalization point $\mu$. 
In the case of an IR finite ghost dressing function, therefore, $u_R(0)$ acquires a non-trivial dependence on the renormalization scale, 
while in the case of an IR divergent ghost dressing function  
the possible $\mu$-dependence is not important, since $u_R(0)=-1$ irrespective of the value of $\mu$ chosen.

Moreover, we introduce the renormalization constant $Z_\lambda$ which relates the bare and renormalized functions, $\lambda_{\mu\nu}^{AB}$ and $\lambda_{\mu\nu,R}^{AB}$, as
\begin{equation}
\lambda_{\mu\nu,R}^{AB}(k, \mu^2)
 =  Z_\lambda(\mu^2,\Lambda^2)  \lambda_{\mu\nu}^{AB}(k,\Lambda^2)
 .
\end{equation}

We find that by virtue of the identities which are the same as (2.17) and (4.1) of \cite{Kondo09a}:
\footnote{
The relation used in \cite{ABP09}
$k^\mu H_{\mu\nu}(p,k)=-i\Gamma_\mu(p,k)$
must be modified in the Gribov-Zwanziger theory. 
}
\begin{subequations}
\begin{align}
 ik^\mu \lambda_{\mu\nu}^{AB}(k)
=&  ik^\mu  \langle   (g \mathscr{A}_\mu \times \mathscr{C})^A (g \mathscr{A}_\nu \times  \bar{\mathscr{C}})^B \rangle_{k}^{\rm m1PI}
= \langle \mathscr{C}^A (g \mathscr{A}_\nu \times \bar{   \mathscr{C}})^B \rangle_k^{\rm 1PI} 
,
\\
 -i \lambda_{\mu\nu}^{AB}(k) k^\nu
=&  -i  \langle   (g \mathscr{A}_\mu \times \mathscr{C})^A (g \mathscr{A}_\nu \times  \bar{\mathscr{C}})^B \rangle_{k}^{\rm m1PI} k^\nu
= \langle (g \mathscr{A}_\mu \times \mathscr{C})^A \bar{   \mathscr{C}}^B \rangle_k^{\rm 1PI}
 ,
\end{align}
\end{subequations}
in the Landau gauge $\lambda_{\mu\nu}^{AB}(k)$ must be renormalized by the same renormalization constant as that for 
 $\langle \mathscr{C}^A (g \mathscr{A}_\nu \times \bar{   \mathscr{C}})^B \rangle_k^{\rm 1PI}$ 
(and $\langle (g \mathscr{A}_\mu \times \mathscr{C})^A \bar{   \mathscr{C}}^B \rangle_k^{\rm 1PI}$) 
which appears in the Schwinger-Dyson equation for the ghost dressing function or the ghost propagator: 
\begin{equation}
 G^{AB}(k)^{-1} := \langle   \mathscr{C}^A \bar{\mathscr{C}}^B \rangle_k^{-1} = -k^2  \delta^{AB} 
-i k^\mu \langle (g \mathscr{A}_\mu \times \mathscr{C})^A \bar{   \mathscr{C}}^B \rangle_k^{\rm 1PI} 
 .
\end{equation}
Therefore, we have
\begin{equation}
 Z_\lambda(\mu^2,\Lambda^2)  =Z_C(\mu^2,\Lambda^2) \tilde{Z}_1^{-1}(\mu^2,\Lambda^2)
 .
\end{equation}

The renormalized quantities must obey
\begin{subequations}
\begin{align}
 1+ u_R(k^2, \mu^2) =& Z_C(\mu^2,\Lambda^2) +   \frac{\tilde{Z}_1(\mu^2,\Lambda^2)}{(D-1)(N^2-1)} \left[ \lambda_{\mu\mu,R}^{AA}(k, \mu^2) - \frac{k^\mu k^\nu}{k^2} \lambda_{\mu\nu,R}^{AA}(k, \mu^2) \right]
  ,
  \label{eq-u-R}
  \\
 w_R(k^2, \mu^2) =&   \frac{ \tilde{Z}_1(\mu^2,\Lambda^2)}{(D-1)(N^2-1)} \left[ D \frac{k^\mu k^\nu}{k^2} \lambda_{\mu\nu,R}^{AA}(k, \mu^2) - \lambda_{\mu\mu,R}^{AA}(k, \mu^2)   \right]
 ,
 \label{eq-w-R}
 \\
  G_R^{-1}(k^2, \mu^2) =& Z_C(\mu^2,\Lambda^2) +   \frac{\tilde{Z}_1(\mu^2,\Lambda^2)}{(N^2-1)} \frac{k^\mu k^\nu}{k^2} \lambda_{\mu\nu,R}^{AA}(k, \mu^2) 
  .
  \label{eq-F-R}
\end{align}
\end{subequations}
If $Z_C$ is determined, we can cast the SD equation for $G_R(k^2):=k^2 G_R(k^2)$ into a manifestly renormalizable form. The same $Z_C$ makes $u_R(k^2)$ finite.

Once the renormalization condition for $G_R$ is chosen to be $G_R(\mu^2)=1$, the values of $u_R(\mu^2)$ is completely determined for its own equation (\ref{eq-u-R}).
It should be noted that the horizon function $\langle  h_R(0) \rangle$ is not a renormalization-group invariant quantity, and hence depends on the  renormalization point $\mu$, just as the Kugo-Ojima parameter $u_R(0)$ depends on the renormalization point $\mu$.

We now study the $\mu$-dependence of (the average of) the horizon function.
\begin{align}
 \langle  h(0) \rangle
=  - \lambda_{\mu\mu}^{AA}(0) -  \Delta_{\mu\mu}^{AA}(0)
 .
\end{align} 
If the renormalization of 
$\Delta_{\mu\nu}^{AB}(k):=  \langle (g \mathscr{A}_\mu \times \mathscr{C})^A \bar{   \mathscr{C}}^C \rangle_k^{\rm 1PI} 
 \langle \mathscr{C}^C \bar{\mathscr{C}}^D \rangle_k 
 \langle \mathscr{C}^D (g \mathscr{A}_\nu \times \bar{   \mathscr{C}})^B \rangle_k^{\rm 1PI} $
 is done according to
$\Delta_{\mu\nu,R}^{AB}(k)=Z_\Delta \Delta_{\mu\nu}^{AB}(k)$ as 
\begin{align}
 \Delta_{\mu\nu,R}^{AB}(k)    
= (\tilde{Z}_1Z_g^{-1} Z_g^{-1} Z_A^{-1} Z_C^{-1})^2 Z_C^{-1}   \Delta_{\mu\nu,R}^{AB}(k)
= \tilde{Z}_1^{-2}  Z_C \Delta_{\mu\nu}^{AB}(k)
:= Z_\Delta \Delta_{\mu\nu}^{AB}(k)
   ,
\end{align}
then the horizon function is renormalized as
\begin{align}
 \langle  h(0) \rangle
=  - \lambda_{\mu\mu}^{AA}(0) -  \Delta_{\mu\mu}^{AA}(0)
=  - Z_C^{-1} [ \tilde{Z}_1 \lambda_{\mu\mu,R}^{AA}(0) +  \tilde{Z}_1^{2} \Delta_{\mu\mu,R}^{AA}(0)]
 .
\end{align}
The Taylor's non-renormalization theorem \cite{Taylor71} says that in the Landau gauge, the renormalization constant  $\tilde{Z}_1$ of the ghost-antighost-gluon vertex is exactly one, $\tilde{Z}_1=1$, see also section 2.2 of \cite{Boucaudetal05}. 

The $\mu$-dependence of the horizon condition would be  obtained from the renormalized horizon condition:
\begin{align}
 \langle  h_R(0) \rangle
= - (N^2-1)  \left\{ Du_R(0) +w_R(0) - G_R(0)[u_R(0)+w_R(0)]^2  \right\}  
 .
\end{align}
This should be compared with the renormalized expression which comes from another form of the bare horizon condition given in a previous paper \cite{Kondo09a}, 
\begin{align}
 \langle  h_R(0) \rangle
= - (N^2-1)  \left\{ (D-1)u_R(0) + G_R(0)[u_R(0)+w_R(0)] \right\}  
 .
\end{align}

The issue of dependence of the Kugo-Ojima function and the horizon function on the renormalization point $\mu$ has been studied in detail by Aguilar, Binosi and  Papavassiliou \cite{ABP09} using a mixed approach, in which the lattice data for the gluon propagator were used as input for their SD equations, see Fig.~9.
According to their results, the renormalized values which are the same as the bare value $u(0)=-0.665 \sim -2/3$ (and $\langle h(0) \rangle = 31.92 \sim (3^2-1)4=32$) are obtained when the renormalization point is chosen at $\mu \sim \sqrt{10.92} {\rm GeV} \sim 3.3 {\rm GeV}$.
\footnote{
The author would like to than Dr. Daniele Binosi for the correspondence on this point.}
  It is possible to translate this result to that at the other renormalization point by using their result. 

The equation for $w_R(k^2)$ remains unchanged, i.e., one simply replaces in the equation for $w(k^2)$ the unrenormalized quantities with the renormalized ones, up to $\tilde{Z}_1$. 
They have shown that $w_R(0)=0$ if both the gluon propagator and the ghost dressing function are IR finite  under the approximation that the form factors are replaced with their tree level values, i.e., bare vertex approximation.
The assumption appears to be a good approximation according to lattice studies \cite{CMM04,IMPSS06}. 
Note that the perturbation theory leads to $w(0) \ne 0$, e.g., $w(k^2)=g^2N/(32\pi^2)$ at one loop, because the free gluon propagator is not IR finite.

The renormalization will be discussed in more detail in a subsequent paper \cite{Kondo09d}.

\section{Remarks }

\subsection{Remark 1: The total derivative term in a choice of the horizon term}

We point out that the result crucially depends on the explicit form of the non-local horizon term adopted.

If the total derivative was neglected  in the Gribov-Zwanziger theory, the horizon term could be rewritten as 
\begin{align}
  \int d^Dx h(x) 
:=&     \int d^Dx \int d^Dy gf^{ABC} \mathscr{A}_\mu^{B}(x) (K^{-1})^{CE}(x,y) gf^{AFE} \mathscr{A}_\mu^{F}(y)
\nonumber\\
?=&    \int d^Dx \int d^Dy D_\mu[\mathscr{A}]^{AC}(x) (K^{-1})^{CE}(x,y) gf^{AFE} \mathscr{A}_\mu^{F}(y)  
\nonumber\\
?=&   \int d^Dx \int d^Dy D_\mu[\mathscr{A}]^{AC}(x) (K^{-1})^{CE}(x,y) D_\mu[\mathscr{A}]^{AE}(y)  
 ,
 \label{GZ3}
\end{align}
which yielded the average of the horizon function:
\begin{align}
 \langle  h(0) \rangle
=  -\lim_{k  \rightarrow 0} \langle   (D_\mu  \mathscr{C})^A (D_\mu   \bar{\mathscr{C}})^A \rangle_{k} 
 ,
\end{align} 
instead of 
$
 \langle  h(0) \rangle
=  - \lim_{k  \rightarrow 0} \langle   (g \mathscr{A}_\mu \times \mathscr{C})^A (g \mathscr{A}_\mu \times  \bar{\mathscr{C}})^A \rangle_{k}
 .
$
Thus, the horizon condition  
$
 \langle  h(0) \rangle
 =(N^2-1)D 
$ for this horizon function defined from this form, i.e., 
\begin{align}
 \langle  h(0) \rangle
=  -\lim_{k  \rightarrow 0} \langle   (D_\mu  \mathscr{C})^A (D_\mu   \bar{\mathscr{C}})^A \rangle_{k} 
= -(N^2-1) 
  \left\{ (D-1)u(0) -1 \right\} =(N^2-1)D
 ,
 \label{hc2}
\end{align} 
 led to the Kugo-Ojima criterion and the divergent ghost dressing function:
 \footnote{
The author would like to thank Drs. David Dudal and Nele Vandersickel for kind discussions on this issue.
}
\begin{equation}
 u(0) = -1 ,  
 \label{KOc}
\end{equation} 
and the divergent ghost dressing function with an imput $w(0)=0$:
\begin{equation}
  G(0) =[1+u(0)+w(0)]^{-1} = w(0)^{-1} = \infty 
 .
\end{equation} 
Thus, if one starts from the horizon term in the last form of (\ref{GZ3}) as done in \cite{Dudal09a} by neglecting the total derivative term in the non-local horizon function,  then one is led to the opposite conclusion to ours. 
This fact that the horizon condition using this definition (\ref{hc2}) is equivalent to the Kugo-Ojima criterion (\ref{KOc}) has already been pointed out and it was checked to what extent the horizon condition holds in the numerical simulation in \cite{NF00}.  
More details on the foundation for the horizon function  are  given in \cite{Zwanziger94}. 

The resulting discrepancy comes from the neglection of the total derivative term in the \textit{non-local} theory. 
In the case of the local theory, the total derivative term can be neglected, since the volume integral is replaced by the surface integral over the surface $S$ far away from the origin by the Gauss theorem:
\begin{align}
 \int_V d^Dx \partial_\mu [f_\mu(x)]
 =& \int_{S=\partial V} d^{D-1} S_\mu(x)  [f_\mu(x)]
 \nonumber\\
 =& \int d\Omega_D |x|^{D-1} n_\mu f_\mu(x)
 =  0
 ,
\end{align}
for a rapidly decreasing and vanishing function at infinity:
$
 n_\mu f_\mu(x) \sim |x|^{-\alpha} \quad (\alpha > D-1) 
 ,
$
where $\Omega_D$ is the solid angle. 

In the case of the non-local theory, however, the total derivative term cannot be neglected, even if the integrand $f(x-y)$ is a rapidly decreasing function in $|x-y|$:
\begin{align}
  \int_V d^Dy \int_V d^Dx \partial_\mu^{x} [f(x-y)g_\mu(y)]
 =&  \int_V d^Dy \int_{S=\partial V} d^{D-1} S_\mu(x) [f(x-y)g_\mu(y)]
 \nonumber\\
 =& \int_{S=\partial V} d^{D-1} S_\mu(x) \int_V d^Dy [f(x-y)g_\mu(y)]
 \ne   0
 ,
\end{align}
because  a contribution from $y$ such that $|x-y|  \sim 0$ always remain due to the existence of an additional integral $\int_V d^Dy$ over the whole volume $V$. 
Even after applying the Gauss theorem, we have still another volume integral and the integrand is not limited to the region far from the origin. 
Thus, if one neglects the total derivative term in the non-local theory, one is led to a theory with different physical content. 

It is a good place to note that, in order to obtain the modified BRST symmetry in \cite{Kondo09b,Sorella09}, integration by parts, i.e., neglection of the total derivative has been used as usual.  However, this is safe, because it is done in the localized Gribov-Zwanziger theory.

\subsection{Remark 2: A contribution from a diagram connected by a single ghost line}

It should be remarked that our result differs from the Zwanziger result \cite{Zwanziger09}.  
The essential difference between his result  and ours comes from the estimation of  $\Delta_{sing}(k):=\Delta_{\mu\mu}^{AA}(k)$.
In fact, if the contribution from this term was neglected,  the average of the horizon function became   equal to (instead of (\ref{r1}))
\begin{equation}
 \langle  h(0) \rangle 
= - \lambda_{\mu\mu}^{AA}(0)
= - (N^2-1)   Du(0)
  ,
 \label{Zwanziger-result}
\end{equation} 
and the horizon condition $\langle  h(0) \rangle=(N^2-1)D$ would become indeed equivalent to the Kugo-Ojima criterion 
\begin{equation}
 u(0)=-1 ,
\end{equation} 
as claimed in \cite{Zwanziger09}. 
However, $\Delta_{sing}(k)$ does not vanish and remains non-zero even after taking the $k=0$ limit, as we have examined in the previous paper \cite{Kondo09a}, which is also the case for $\gamma \ne 0$.

\section*{Acknowledgments}
The author would like to thank Michael M\"uller-Preussker, Marc Wagner, Hilmar Forkel, Dietmar Ebert, Tereza Mendes, Frieder Lenz, Michael Thies, Falk Bruckmann, Andreas Sch\"afer, Gunnar Bali, Jacques Bloch, Hugo Reinhardt, Peter Watson, Olivier Pene, Philippe Boucaud, Jose Rodriguez-Quintero, Henri Verschelde, David Dudal and Nele Vandersickel for helpful discussions, and Taichiro Kugo, Hideo Nakajima, Sadataka Furui, Danielle Binosi and Andre Sternbeck for correspondences. 
He is grateful to High Energy Physics Theory Group and Theoretical Hadron Physics Group in the University of Tokyo, especially, Prof. Tetsuo Hatsuda for kind hospitality extended to him on sabbatical leave.
This work is financially supported by the exchange program between JSPS and DAAD (from 14 June to 4 July 2009), 
and Grant-in-Aid for Scientific Research (C) 21540256 from Japan Society for the Promotion of Science
(JSPS).

\appendix
\section{Symmetry of exchanging ghost and antighost}

In the Landau gauge, $\partial_\mu \mathscr{A}_\mu^A(x)=0$, the equations of motion for the ghost field and the antighost field have the same form
\begin{equation}
\partial_\mu D_\mu[\mathscr{A}] \mathscr{C}^A = 0,
\quad \partial_\mu D_\mu[\mathscr{A}] \bar{\mathscr{C}}^A = 0 
 , 
\end{equation}
This suggests that there is a symmetry under the exchange of ghost and antighost fields. 
In fact, the GZ action is invariant under the \textit{FP conjugate transformation}: 
\begin{equation}
  \mathscr{C}^A \rightarrow \bar{\mathscr{C}}^A ,
\quad
  \bar{\mathscr{C}}^A \rightarrow - \mathscr{C}^A ,
\quad
  \mathscr{B}^A \rightarrow - \bar{\mathscr{B}}^A ,
  \label{FPconjugation}
\end{equation}
where other fields are unchanged and $\bar{\mathscr{B}}$ is defined by
\begin{equation}
  \mathscr{B}^A   + \bar{\mathscr{B}}^A
=   g (i\bar{\mathscr{C}}  \times \mathscr{C})^A .
\end{equation}
This is easily seen by rewriting the GF+FP term as
\begin{align}
 S_{\rm GF+FP}  :=& \int d^Dx  \left\{ \mathscr{B} \cdot \partial_\mu \mathscr{A}_\mu 
+i \bar {\mathscr{C}} \cdot \partial_\mu D_\mu \mathscr{C} \right\} 
:=& \int d^Dx  \left\{ - \bar{\mathscr{B}} \cdot \partial_\mu \mathscr{A}_\mu 
-   \mathscr{C} \cdot \partial_\mu D_\mu i\bar {\mathscr{C}} \right\} 
  .
\end{align}

\section{Another derivation of eq.(\ref{dDCC})}

We work in the Minkowski formulation. 
In the canonical quantization, the relation (\ref{dDCC}), i.e., 
\begin{align}
 \partial^\mu_{x} \langle 0| T (D_\mu \mathscr{C})^A(x)  \bar{\mathscr{C}}^B(y) |0 \rangle 
= \delta^{AB} \delta^D(x-y)  
 ,
\end{align}
is derived from the equation of motion for the ghost field
$\partial ^{\mu}D_{\mu}\mathscr{C}(x)=0$ 
and the equal-time canonical anti-commutation relation$
\{\bar{\mathscr{C}}^A(x),i(D_0\mathscr{C})^B(y)\}_{x_0=y_0}=i\delta ^{AB}\delta ^{(D-1)}(\mbox{\boldmath $x$}-\mbox{\boldmath $y$})
$.
In fact, the equation of motion
 \begin{align}
0=   \frac{\partial \mathscr{L}_{\rm GZ}}{\partial \bar{\mathscr{C}}^A(x)}
 =&  -i \partial^\mu (D_\mu \mathscr{C})^A(x)   
 ,
\end{align}
and the equal-time canonical anti-commutation relation 
\begin{align}
i \delta^{AB} \delta^{D-1}(x-y) 
=  \{ \Pi^A_{\bar{\mathscr{C}}}(x),  \bar{\mathscr{C}}^B(y) \}_{x_0=y_0}
 ,
\end{align}
implies
\begin{align}
 & \partial^\mu_{x} \langle 0| T (D_\mu \mathscr{C})^A(x)  \bar{\mathscr{C}}^B(y) |0 \rangle 
\nonumber\\
 =& \delta(x_0-y_0)  \langle 0| \{ (D_0 \mathscr{C})^A(x)  \bar{\mathscr{C}}^B(y) \} |0 \rangle 
 +  \langle 0| T \partial^\mu (D_\mu \mathscr{C})^A(x)  \bar{\mathscr{C}}^B(y) |0 \rangle 
\nonumber\\
 =& \delta^{AB} \delta^D(x-y)  
 ,
\end{align}
where we have used
\begin{align}
 \Pi^A_{\bar{\mathscr{C}}}(x) 
=   \frac{\partial \mathscr{L}_{\rm GZ}}{\partial \partial_0 \bar{\mathscr{C}}^A(x)} 
=   i(D_0 \mathscr{C})^A(x) 
 ,
\end{align}
and the time-ordered product (T-product) for the anti-commuting field defined by
\begin{equation}
 TA(x)B(y) := \theta(x_0-y_0) A(x)B(y) - \theta(y_0-x_0) B(y) A(x) .
\end{equation}


\end{document}